\documentclass[letterpaper]{article}

\usepackage[T1]{fontenc}

\usepackage{geometry}
\usepackage{setspace}

\usepackage[style = chem-acs]{biblatex}
\usepackage{graphicx}
\usepackage{float}
\usepackage{float}
\newfloat{scheme}{htbp}{los}
\floatname{scheme}{Scheme}
\floatname{chart}{Chart}
\newfloat{graph}{htbp}{loh}
\usepackage{amssymb}

\usepackage{chemformula} 
\usepackage[version = 4]{mhchem} 

\usepackage{authblk}
\author[1]{Sajjan Sheoran*}
\author[1]{Luke Keenan}
\author[1]{Declan Nell}
\author[1]{Stefano Sanvito$^{\dagger}$}
\affil[1]{School of Physics and CRANN Institute, Trinity College, Dublin 2, Ireland}

\title{Negative Differential Resistance and Ultra-High TMR in Altermagnetic Tunnel Junctions}
\date{*Email: sajjans@tcd.ie; $\dagger$Email: sanvitos@tcd.ie}

\begin{document}

\maketitle

\begin{abstract}
Altermagnets can replace ferromagnets in tunnel junctions, yielding large tunneling 
magnetoresistance, ultrafast switching, and low-power functionality. While most studies 
explore the linear-response regime, interesting features emerge at finite bias, where the peculiar 
electronic structure of altermagnets gives rise to complex non-linear behaviour. Using 
non-equilibrium Green's functions implemented with density functional theory, we predict that a large low-bias
negative differential resistance can be observed in an altermagnetic tunnel junction. 
Our proposed junction incorporates the orbital-ordered altermagnet KV$_2$Se$_2$O, whose 
quasi-2D Fermi surface plays a crucial role in realizing the negative differential resistance. Upon 
the application of a finite bias voltage, the current in the parallel configuration first increases 
sharply and then decreases, to be almost completely suppressed at around 0.14~V. At the same 
time, the antiparallel configuration displays a monotonic current-voltage curve. This behaviour, in 
addition to the negative differential resistance, supports a large tunneling magnetoresistance with 
sign inversion at 0.13~V. Our results suggest that altermagnetic tunnel junctions can be 
used as components in applications requiring strong non-linear response at low bias. 
\end{abstract}


Altermagnets (AMs) have emerged as a fundamentally novel class of collinear magnetic materials, 
combining compensated magnetic order in real space with broken time-reversal symmetry in momentum space~\cite{vsmejkal2022beyond,yuan2021prediction,hayami2020bottom,reichlova2024observation, sheoran2025spontaneous}. 
The exchange-driven eV-scale spin-splitting of AMs opens a route to spintronic functionalities conventionally 
associated with ferromagnets (FMs), including spin-polarized transport~\cite{gonzalez2021efficient}, anomalous 
Hall effect~\cite{gonzalez2023spontaneous}, chiral magnons~\cite{vsmejkal2023chiral,liu2024chiral}, and 
magnetoresistive response~\cite{gonzalez2024anisotropic,zhang2025theory}. In particular, AMs are highly 
promising as a platform for magnetic tunnel junctions owing to negligible stray fields and ultrafast switching 
enabled by the vanishing net magnetization~\cite{xu2025giant,noh2025tunneling}. In contrast to conventional 
ferromagnetic tunnel junctions (FMTJs), where transport is commonly associated to spin 
polarization~\cite{halder2024half, halder2026spin} and symmetry filtering~\cite{butler2001spin,mathon2001theory,mavropoulos2000complex,caffrey2012coexistance, nell2025effect}, 
in altermagnetic tunnel junctions (AMTJs) this exhibits a stronger dependence on the momentum-resolved 
electronic structure of the leads~\cite{vsmejkal2022giant,noh2025tunneling}. Indeed, AMTJs have been predicted 
to exhibit extremely high linear-response tunneling magnetoresistance (TMR), reaching $\sim$$10^8$, due to 
transverse-momentum ($\boldsymbol{k}_{\parallel}$) matching across the barrier in the parallel (P) configuration 
and its strong suppression in the antiparallel (AP) one~\cite{shao2021spin, vsmejkal2022giant, jiang2023prediction,Chi2024PRApp, Liu2024PRB, Chi2025PRApp, Sun2025PRB, Yang2025PRB, yang2026altermagnetic, he2026tunnel}.

Beyond the linear-response, finite-bias non-equilibrium transport in AMTJs remains largely 
unexplored. Among the possible non-linear tunneling phenomena, negative differential resistance (NDR), 
the peculiar decrease of the current with increasing bias, is of particular interest for high-frequency electronics, 
multi-valued logic, static random-access memory, and magnetic random-access 
memory~\cite{chen1999large,Britnell2013resonant,Zhang2025toward}. NDR has been extensively investigated 
in Esaki and resonant-tunneling diodes, and both molecular and van der Waals tunnel 
junctions~\cite{Esaki1958new,Britnell2013resonant,Dalgleish2006interface,Shim2016phosphorene}. In such systems, 
the NDR is most commonly attributed to quenching resonant transmission channels, the evolution of quantum-well 
or interface states, or bias-induced reduction in spin 
polarization~\cite{Britnell2013resonant,Dalgleish2006interface,Shim2016phosphorene,burg2017coherent,saha2012magnetoresistance}.

As we will demonstrate, AMTJs appear as natural candidates to implement NDR, which can be driven not only by 
energy-level alignment, but also by the features of the $\boldsymbol{k}_{\parallel}$-dependent transmission. 
In an AMTJ where $d$-wave AM electrodes have a quasi-2D Fermi surface, a finite bias voltage, V$_b$, suppresses 
coherent tunneling across the barrier. Using density functional theory (DFT)~\cite{kohn1965self} non-equilibrium Green-function (NEGF) 
simulations~\cite{rungger2020non}, we report a pronounced NDR in AMTJs based on the recently synthesized family of vanadium oxychalcogenides, 
$A$V$_2Q_2$O ($A$=K, Rb, Cs; $Q$=S, Se, Te)~\cite{jiang2025metallic,zhang2025crystal,liu2025physical,yang2026visualizingspinpolarizationaltermagnetkv2se2o,wang2025atomic,fu2025atomic,thapa2026altermagnetism}. In the P configuration, the current, $I$, of the $\mathrm{KV_2Se_2O|MgO|KV_2Se_2O}$ 
junction peaks at $\sim$$0.63$~nA for V$_b\sim0.07$ V and then rapidly decreases, becoming nearly quenched 
for $V_b \gtrsim 0.14$~V. This leads to a $\Lambda$-type NDR. In contrast, the AP configuration displays a smooth 
monotonic $I$-$V_b$ curve, resulting in a finite-bias order-of-magnitude TMR drop followed by a sign inversion.
The NDR behavior applies to the large class of crystal-symmetry-paired AMs with similar quasi-2D open Fermi 
sheets weakly dispersed along $k_z$~\cite{jiang2025metallic,zhang2025crystal,liu2025physical,thapa2026altermagnetism,hu2025catalog}. 

To understand how the NDR arises, consider the minimal \(d\)-wave tight-binding Hamiltonian~\cite{vsmejkal2022giant},
\begin{equation}\label{Eq1}
\begin{aligned}
H(\boldsymbol{k})={}&\left[2t\left(\cos k_x+\cos k_y\right)-\mu\right]\hat{\sigma}_0+ 2t_J\left(\cos k_x-\cos k_y\right)\hat{\sigma}_z\:,
\end{aligned}
\end{equation}
where \(t\) denotes the nearest-neighbor kinetic hopping amplitude on the square lattice, while \(t_J\) describes 
the spin-momentum anisotropy. Here, \(\mu\) is the chemical potential, and \(\hat{\sigma}_0\) and \(\hat{\sigma}_z\) 
are the identity and the $z$ Pauli matrix, respectively, acting in spin space. Note that we have neglected any \( k_z\)-dependence 
in the Hamiltonian, reflecting the quasi-2D nature of the material with weak coupling along the $z$-direction. Furthermore, 
we consider the maximally spin-anisotropic regime, \(|t| \simeq |t_J|\), in which spin-up and spin-down electrons hop 
predominantly along mutually perpendicular directions. As a consequence, the Fermi surface is quasi-2D, consisting of 
quasi-1D flat sheet-like segments in three dimensions, as illustrated in Fig.~1(a) [see also Section S1 of the Supporting 
Information (SI)]. Its projection onto the $\boldsymbol{k}_{\parallel}$ plane therefore appears as nearly straight lines, rather 
than the closed circular or elliptical contours familiar for more isotropic systems~\cite{vsmejkal2022beyond}. Such 
direction-dependent hopping and Fermi surfaces can arise in  orbital-ordered AMs stabilized by electronic 
correlations~\cite{leeb2024spontaneous,kaushal2025altermagnetism,li2026altermagnetic} and AMs with X-type 
cross-chain opposite spin sublattices~\cite{zhang2025x,wang2025high}.
\begin{figure}
    \centering
    \includegraphics[width=0.5\linewidth]{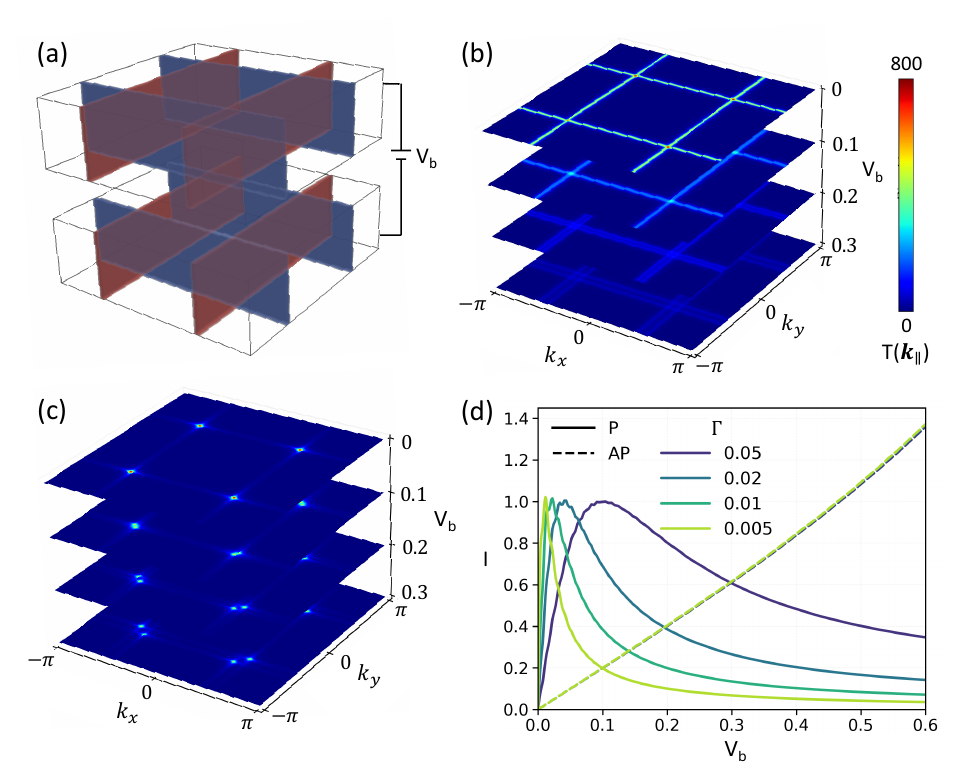}
    \caption{Simple model analysis. In panel (a) we illustrate the P configuration of the AMTJ with included the spin-resolved 
    Fermi surfaces described in Eq.~(\ref{Eq1}) ($t=t_J=1$ and $\mu=0$). The red (blue) surfaces are the spin-up (spin-down) 
    Fermi surfaces. In the AP configuration, the N\'eel vector of one altermagnetic lead is switched relative to the other. Panels
    (b) and (c) show the $\boldsymbol{k}_{\parallel}$-resolved transmission for the P and AP configurations at equilibrium 
    Fermi energy, respectively. In (d) 
    we present the $I$-$V$ curve for the P and AP configurations for different spectral broadening, $\Gamma$.}
    \label{fig:1}
\end{figure}

\begin{figure}
    \centering
    \includegraphics[width=1.00\linewidth]{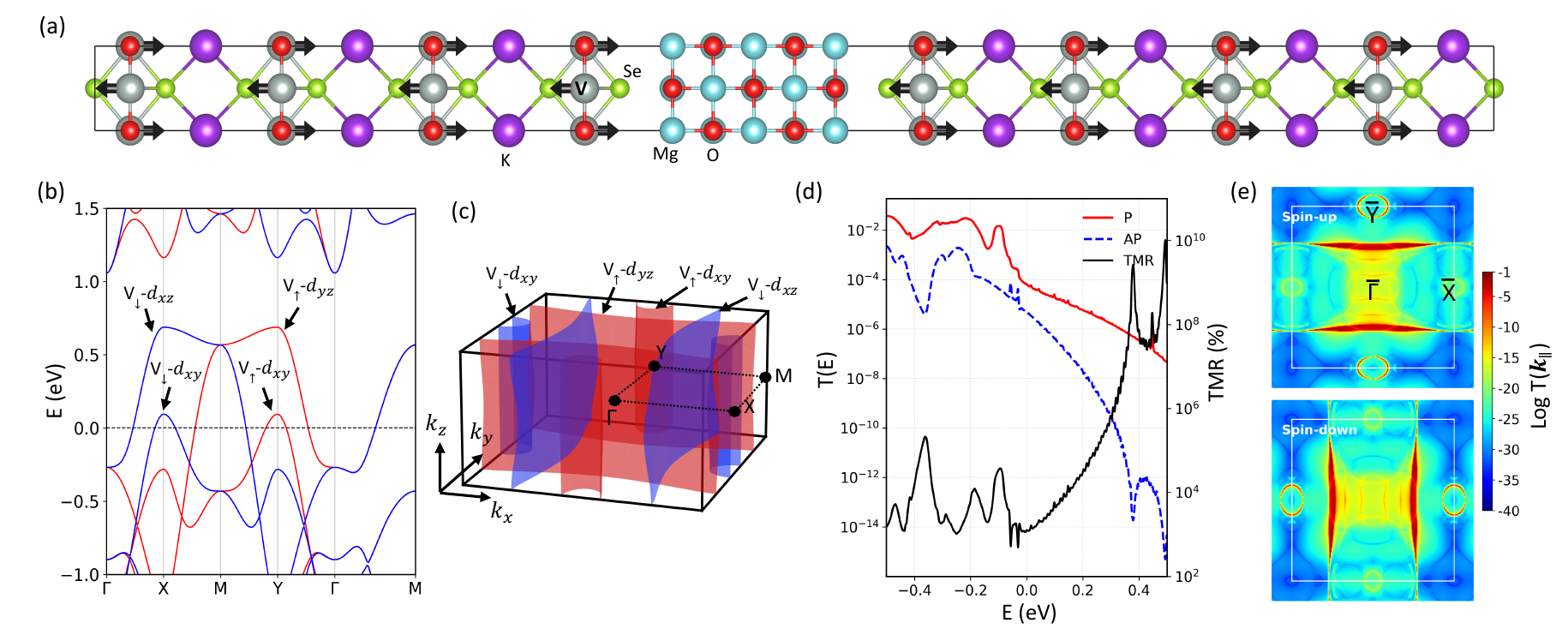}
    \caption{Zero-bias transport properties of the $\mathrm{KV_2Se_2O|MgO|KV_2Se_2O}$ junction. (a) the atomistic structure 
    of the scattering region in the P configuration of the N\'eel vectors. This is connected to two semi-infinite $\mathrm{KV_2Se_2O}$ 
    leads on both sides. The AP configuration is obtained by reversing the N\'eel vector of the right lead. (b) Spin-polarized 
    band structure and (c) Fermi surface of $\mathrm{KV_2Se_2O}$. Red surfaces are for spin up, while the blue ones for
    spin down. (d) Transmission and TMR as functions of energy at zero bias ($E_\mathrm{F}=0$). 
    (e) $\boldsymbol{k}_{\parallel}$-resolved transmission for the spin-up and spin-down 
    states at $E_\mathrm{F}$ in the P configuration. See Fig. S5 in the SI for the $\boldsymbol{k}_{\parallel}$-resolved 
    transmission of the AP configuration. The white square in (e) is 
    $\boldsymbol{k}_{\parallel}$-projected Brillouin zone.}
    \label{fig:2}
\end{figure}

We then consider a vertical AMTJ setup [see Fig.~\ref{fig:1}(a)] and compute the Landauer--B\"uttiker tunneling 
current at finite bias~\cite{datta1997electronic} (see Section S1 of the SI for details). Within the two-spin-fluid 
picture~\cite{mott1936electrical}, the spin-$\sigma$ transmission coefficient, \(T^{\sigma}(E)\), 
is proportional to the spectral overlap of the two AM electrodes~\cite{sanvito2005ab}. 
For energy- and momentum-independent quasi-particle broadening, $\Gamma$, the spin-resolved spectral 
function is a Lorentzian~\cite{mahan2013many},
\begin{equation}
A_{\sigma}(\boldsymbol{k}_{\parallel},E)=\frac{2\Gamma}{\left[E-\mathcal{E}_{\sigma}(\boldsymbol{k}_{\parallel})\right]^2+\Gamma^2},
\end{equation}
where \(\mathcal{E}_{\sigma}(\boldsymbol{k}_{\parallel})\) is the energy dispersion of Eq.~(\ref{Eq1}). The 
\(\boldsymbol{k}_{\parallel}\)-dependent \(T^{\sigma}(E)\) at the equilibrium Fermi energy ($E_\textrm{F}$) for the 
P and AP alignments of the electrodes are shown
respectively in Fig.~\ref{fig:1}(b) and Fig.~\ref{fig:1}(c). Here we define the junction configuration 
according to the relative orientation of the N\'eel vectors of the electrodes [see Fig.~\ref{fig:2}(a)]. In the 
P configuration, the large transmission at zero bias originates from the complete overlap of the projected 
Fermi sheets over the $\boldsymbol{k}_{\parallel}$ plane. As $V_b$ is applied, the overlap is reduced, since 
the Fermi surfaces of the two electrodes get offset with respect to each other, with a consequent suppression 
of the transmission. In the AP configuration, for the same spin the projected Fermi sheets of the two 
electrodes are mutually orthogonal, so that their overlap remains nearly unchanged under bias. 
Thus, as the bias window opens, the P transmission is progressively suppressed, while the AP
one remains roughly constant. As a consequence, the $I$-$V_b$ curve for the P state displays
an NDR, with the peak current having an almost constant amplitude, however the bias value at which 
it occurs scales linearly with $\Gamma$ [Fig.~\ref{fig:1}(d)]. In contrast, the AP configuration has a trivial monotonic $I$-$V_b$.

The physics of our simple toy model can be realised in an AMTJ with $\mathrm{KV_2Se_2O}$ electrodes
and MgO non-magnetic spacer [see Fig.~\ref{fig:2}(a)].  $\mathrm{KV_2Se_2O}$-based AMTJs have 
recently gained attention due to their predicted high TMR~\cite{yang2026altermagnetic,he2026tunnel,liu2026intrinsic,Yang2025PRB}. 
$\mathrm{KV_2Se_2O}$ belongs to the inverse-Lieb-lattice family and has recently been established experimentally 
as a metallic room-temperature $d$-wave
altermagnet~\cite{jiang2025metallic,zhang2025crystal,liu2025physical,thapa2026altermagnetism,lai2025d}. MgO is 
the non-magnetic spacer of choice because of its wide-gap insulating character, and closely matched lattice symmetry 
and lattice constant to $\mathrm{KV_2Se_2O}$. It is also well-established as tunnel barrier in magnetic junctions, 
such as $\mathrm{Fe|MgO|Fe}$~\cite{nell2025effect}, $\mathrm{Ta|MgO|Mn_3Sn}$~\cite{xie2022magnetization}, 
$\mathrm{RuO_2|MgO|RuO_2}$~\cite{xu2025giant}, and $\mathrm{CoFeB|MgO|CoFeB}$~\cite{ikeda2010perpendicular}. 
In our transport setup the scattering region comprises four KV$_2$Se$_2$O unit cells at each side of the 5-monolayer-thick
MgO barrier to fully converge the electrostatic potential. The system is driven out-of-equilibrium by shifting the local
chemical potential of the right (left) lead by $\mu_\mathrm{R}=E_\mathrm{F}- eV_b/2$ 
($\mu_\mathrm{L}=E_\mathrm{F}+ eV_b/2$), where $e$ is the 
electronic charge. Transport across the junction is computed using the NEGF formalism as implemented in 
the \textsc{Smeagol}~\cite{rocha2006spin,PhysRevB.78.035407} package, interfaced with the 
\textsc{Siesta}~\cite{soler2002siesta} DFT engine (see SI Section S2 for simulations details).

$\mathrm{KV_2Se_2O}$ has nontrivial spin point group $^24/^1m^1m^2m$, where the two opposite-spin 
sublattices are related by the $[C_2||C_{4z}]$ operation, made of a two-fold spin rotation $C_2$ and a four-fold 
real-space rotation $C_{4z}$. These symmetry fingerprints directly reflect into the band structure, where the spin polarization 
is reversed between the $\mathrm{\Gamma}$-X and $\Gamma$-Y directions, leading to a $d_{x^2-y^2}$-wave spin-splitting, 
Fig.~\ref{fig:2}(b). Four bands, predominantly derived from $V_\downarrow$-$d_{xz}$, $V_\uparrow$-$d_{yz}$, 
$V_\uparrow$-$d_{xy}$ and $V_\downarrow$-$d_{xy}$ orbitals, cross the Fermi level. The $V_\uparrow$-$d_{xy}$ 
and $V_\downarrow$-$d_{xy}$ states hop equivalently along the in-plane directions, giving rise to nearly cylindrical 
Fermi surfaces extending along $k_z$. By contrast, the $V_\downarrow$-$d_{xz}$ and $V_\uparrow$-$d_{yz}$ states 
exhibit strongly anisotropic hopping and thus realize the limit of maximal spin anisotropy. In particular, the $d_{xz}$ and 
$d_{yz}$ orbitals disperse predominantly along the $x$ and $y$ directions, respectively. As a result, these states constitute 
the sheetlike Fermi surfaces of interest, with weak dispersion in the $yz$ and $xz$ planes [Fig.~\ref{fig:2}(c)]. These 
orbital-ordered AM features place $\mathrm{KV_2Se_2O}$ among the promising candidates for observing NDR.

We first compute the transmission coefficients as a function of energy for the $\mathrm{KV_2Se_2O|MgO|KV_2Se_2O}$ 
at zero bias [see Fig.~\ref{fig:2}(d)]. The transmission broadly follows the KV$_2$Se$_2$O density of states (DOS) 
[see Fig. S4 of the SI], namely it decreases with increasing energy. In the P configuration 
$T$ is nearly two orders of magnitude larger than in the AP one, because of the $\boldsymbol{k}_{\parallel}$-matching of 
the spin-resolved transmission channels in the two leads, as noted for our simple model. Consistently, the 
$\boldsymbol{k}_{\parallel}$-resolved transmission in Fig.~\ref{fig:2}(e) follows the Fermi surface projection onto the 
$\boldsymbol{k}_{\parallel}$-plane [see Fig. S6 of SI for the $\boldsymbol{k}_{\parallel}$-projected Fermi surface]. The AP 
configuration suppresses the $\boldsymbol{k}_{\parallel}$-matching of the spin-resolved 
channels in the two leads, as the same-spin Fermi sheets are mutually perpendicular in momentum space. As a result, 
the AP transmission is determined by four isolated $\boldsymbol{k}_{\parallel}$ points, where the same-spin Fermi sheets 
of the two leads intersect [see Fig. S5 of the SI]. The ``optimistic'' TMR ratio, defined as 
\(\frac{T_{\mathrm{P}}-T_{\mathrm{AP}}}{\min\!\left(T_{\mathrm{P}},T_{\mathrm{AP}}\right)}\) ($T_\alpha$ is the total transmission of the $\alpha$ configuration), 
is then $\sim 10^3$\% at $E_\mathrm{F}$ and increases up to $\sim 10^{10}$\% at higher energies. The spin-up and spin-down 
$\boldsymbol{k}_{\parallel}$-resolved transmission are related by $C_4$ altermagnetic symmetry and therefore have 
equal magnitude, Fig.~\ref{fig:2}(e). Therefore, the net spin polarization, defined as 
$\frac{T_{\uparrow} (E_\textrm{F})-T_{\downarrow}(E_\textrm{F})}{T_{\uparrow}(E_\textrm{F})+T_{\downarrow}(E_\textrm{F})}$, 
is zero in AMTJs, where the TMR is dictated by the AM spin anisotropy rather than the net spin polarization.
\begin{figure}[h]
    \centering
    \includegraphics[width=0.5\linewidth]{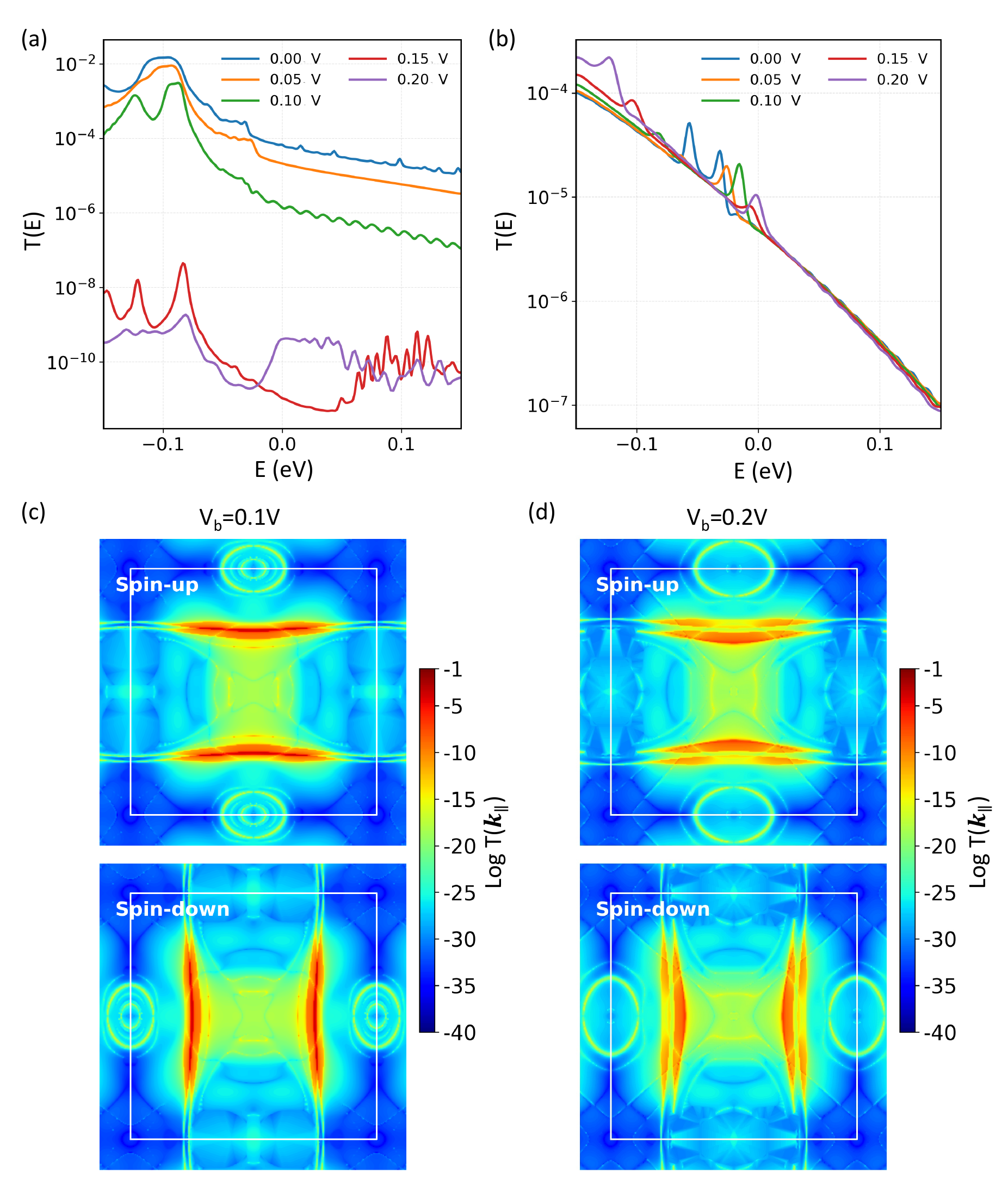}
    \caption{Bias-dependent transmission properties of the $\mathrm{KV_2Se_2O|MgO|KV_2Se_2O}$ junction. 
    The transmission coefficients for the (a) P and (b) AP configuration are presented as a function of energy for 
    different voltages. In (c) and (d) we show the $\boldsymbol{k}_{\parallel}$- and spin-resolved transmission 
    coefficient at $E_{\textrm{F}}$ in the P configuration at a finite bias of $0.1$~V and $0.2$~V, respectively. 
    Note that the heat map is on a logarithmic scale.}
    \label{fig:3}
\end{figure}

Next, we switch on a finite voltage and compute self-consistently the potential drop across the junction. The 
transmission at finite bias is shown in Figs.~\ref{fig:3}(a) and \ref{fig:3}(b). The most
striking feature is that $T_\mathrm{P}$ gets suppressed by orders of magnitude as the bias increases,
and collapses to zero beyond a threshold voltage of 0.14~V. In contrast, the AP transmission remains nearly unchanged 
as a function of bias. A voltage displaces the electrochemical potentials of the left and right leads by $\pm eV_b/2$, 
causing the corresponding Fermi sheets to drift in momentum space. Owing to the strongly direction-dependent hopping, 
the $V_{\uparrow}$-$d_{xz}$ and $V_{\uparrow}$-$d_{yz}$ states shift predominantly along the $x$ and $y$ directions, 
respectively [Fig.~\ref{fig:2}(c)]. As a result, the Fermi sheets largely preserve their mutual orthogonality under bias, while 
their positions move relative to the $\Gamma$ point [see Fig. S6 in the SI for constant energy surfaces]. For an upward 
shift of the chemical potential, they drift away from $\Gamma$, whereas for a downward shift, they move toward $\Gamma$. 
For the P configuration, the large transmission at zero bias originates from the complete $\boldsymbol{k}_{\parallel}$-space 
overlap between the spin-polarized channels of the two leads. At finite bias, however, the Fermi sheets of the left and right 
leads drift in opposite directions, progressively reducing the overlap [Figs.~\ref{fig:3}(c) and \ref{fig:3}(d)]. The characteristic 
threshold bias is determined by the spectral broadening of the Fermi sheets and the Fermi velocity. Once the bias-induced 
separation exceeds this broadening, the transmission vanishes [see Fig.~\ref{fig:2}(d)]. A smaller contribution to the transmission 
also originates from the Fermi tubes associated with the $V_{\uparrow/\downarrow}$-$d_{xy}$ states around the X and Y points. 
These expand (contract) for downward (upward) shifts of the chemical potential, but contribute
little to the total transmission. 

In contrast, in the AP configuration, the same-spin Fermi sheets of the two leads are mutually orthogonal, and at 
finite bias, the four transmission hotspots split into eight (four each for spin-up and spin-down), corresponding to the sheets 
intersection positions. Note that the magnitude of the transmission at the hotspots depends little on the voltage, although their 
position in $\boldsymbol{k}_{\parallel}$-space changes with bias [see Fig. S5 of the SI]. Consequently, the total AP transmission 
remains nearly unchanged with bias [Fig.~\ref{fig:3}(b)].
\begin{figure}[h]
    \centering
    \includegraphics[width=0.6\linewidth]{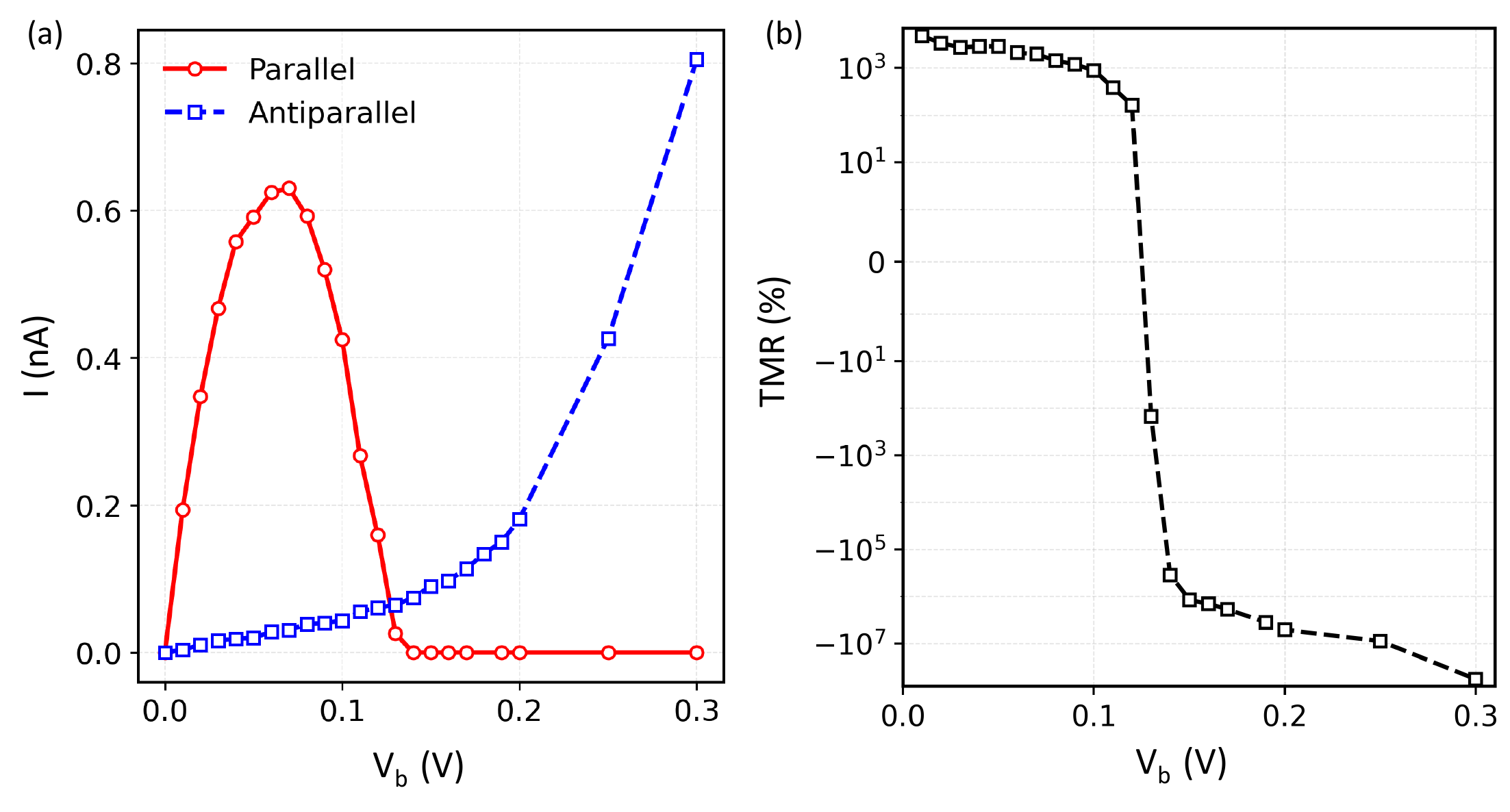}
    \caption{(a) $I$-$V_b$ characteristics of the $\mathrm{KV_2Se_2O|MgO|KV_2Se_2O}$ junction for the P 
    and AP configurations and associated (b) bias-dependent ``optimistic'' TMR, calculated as 
    {\(\frac{I_{\mathrm{P}}-I_{\mathrm{AP}}}{\min\!\left(I_{\mathrm{P}},I_{\mathrm{AP}}\right)}\) and plotted on 
    a symmetric logarithmic scale}.}
    \label{fig:4}
\end{figure}

The currents for the P and AP configuration, respectively $I_{\mathrm P}$ and $I_{\mathrm{AP}}$, 
are computed by integrating the bias-dependent transmission coefficients over the bias window. As 
$V_b$ increases, $I_{\mathrm P}$ initially rises due to the opening of the transport window, but 
subsequently decreases, as a result of the strong suppression of $T$ [Fig.~\ref{fig:4}(a)]. The current 
reaches a maximum value of $0.63$~nA at $V_b=0.07$~V. Then, beyond the threshold bias $V_b\gtrsim 0.14$~V, 
$I_{\mathrm P}$ almost vanishes. This behaviour defines a $\Lambda$-type NDR, where the current 
has a peak and then descends to zero. A $\Lambda$-type NDR is preferred over the N-~\cite{Esaki1958new} 
or S-type~\cite{simpson2002analysis} in conventional two-terminal diodes, whose efficiency is measured 
in terms of peak-to-valley current ratios.  At significantly higher bias voltages, $V_b\gtrsim 0.50$~V, $I_{\mathrm P}$ 
begins to increase again, as states lying approximately $0.25$~eV below $E_\mathrm{F}$ enter the transport 
window [see Fig.~S7 of the SI]. 
In contrast, as expected, $I_{\mathrm{AP}}$ has a monotonic behaviour and crosses over $I_{\mathrm{P}}$
at around $V_b\gtrsim 0.13$~V. As a consequence, the optimistic finite-bias TMR, defined using currents as 
$\mathrm{TMR}\equiv \frac{I_{\mathrm{P}}-I_{\mathrm{AP}}}{\min\!\left(I_{\mathrm{P}},I_{\mathrm{AP}}\right)}$,
is about $4\times10^3$\% at zero-bias and decreases sharply to vanish at $V_b\gtrsim 0.13$~V. However, beyond 
$V_b= 0.13$~V it changes sign and rapidly increases in magnitude again, reaching up to $-5\times10^7$\% at $V_b\sim0.3$~V.
Note that $I_{\mathrm{P}}$ and $I_{\mathrm{AP}}$ remain spin neutral as the spin-up and spin-down 
channels contribute equally to the total transport. 

Let us quickly emphasize the difference between the NDR observed here for AMTJs and that  
sometimes found in conventional magnetic tunnel junctions. When ferromagnetic electrodes are
used, the transmission in the P and AP configurations can be explained through the Julliere's 
formula~\cite{julliere1975tunneling}, which is strictly appropriate only in the case of amorphous 
barriers, but helps the understanding here. According to Julliere, we have 
$T_\mathrm{P}\propto D_{1\uparrow}D_{2\uparrow}+D_{1\downarrow}D_{2\downarrow}$ 
and $T_{\mathrm{AP}}\propto D_{1\uparrow}D_{2\downarrow}+D_{1\downarrow}D_{2\uparrow}$, where 
$D_{\alpha\sigma}$ is the DOS of lead $\alpha=\mathrm{L, R}$ for spin $\sigma=\uparrow, \downarrow$.
The NDR, if any, then arises when some energy-narrow features of the DOS sweeps across the bias window (see,
for example, Refs.~\cite{zhao2021electric,saha2012magnetoresistance, PhysRevB.64.075420, zimbovskaya2008negative}). 
In the case of AMTJs, the energy-dependent DOS remains broadly unchanged under bias, but the NDR 
originates from sharp features of the DOS in $\boldsymbol{k}_{\parallel}$ space. 

In summary, our research establishes a novel robust mechanism for engineering NDRs. This occurs in AMTJs 
with electrodes having quasi-2D flat Fermi surfaces. The fundamental relation between the NDR and the spin 
anisotropy is studied first through transport simulations based on an effective model Hamiltonian, which helps 
in identifying the main mechanism for the effect. Then, we propose an actual practical device, 
KV$_2$Se$_2$O|MgO|KV$_2$Se$_2$O, where the NDR can be detected. This uses the prototypical $d$-wave 
AM, KV$_2$Se$_2$O. A $\Lambda$-type NDR is observed with the associated large TMR changing sign at a
critical voltage close to that where the parallel current vanishes. While there is experimental ambiguity on whether 
the ground state of KV$_2$Se$_2$O is AM or a conventional G-type anti-ferromagnet (also defined as anti-altermagnetism), 
AM in KV$_2$Se$_2$O is confirmed by spin and angle-resolved photoemission spectroscopy~\cite{jiang2025metallic}, 
and by spin-selective tunneling~\cite{yang2026visualizingspinpolarizationaltermagnetkv2se2o,wang2025atomic,fu2025atomic}. 
This suggests that our proposed device may actually be fabricated in the lab. Importantly, our findings are also relevant to the other AMs with similar quasi-2D Fermi surfaces in $k$-space. These includes, for example, other vanadium
oxychalcogenides, Rb$_{1-\delta}$V$_2$Te$_2$O~\cite{zhang2025crystal}, CsV$_2$Se$_2$O~\cite{fu2025atomic}, 
Ruddlesden-Popper chromates, SrCrO$_3$~\cite{meier2026net}, organic compound 
$\kappa$-(ET)$_2$Cu[N(CN)$_2$]Cl~\cite{yu2025altermagnetism}, and layered oxides nickelates, LaNiO$_3$~\cite{maznichenko2024fragile}. The AMTJ can be further optimized for NDR by selecting a non-magnetic spacer with a slow decay rate for the flat Fermi-sheet states. 

\section*{Acknowledgements}
The research conducted in this publication was funded through the Government of Ireland Postdoctoral Fellowship by Research Ireland under grant number GOIPD/2025/1092. Computational resources were provided by Trinity College Dublin Research IT.

\section*{Supporting information}

Supporting information is available:
\begin{itemize}
  \item SI.pdf: Supporting Information provides detailed calculations and supplement figures referred in the main manuscript.
\end{itemize}


\section{TOC Graphic}
\begin{minipage}[b][1.75in]{3.25in}
  \sffamily
  \frenchspacing
   \includegraphics{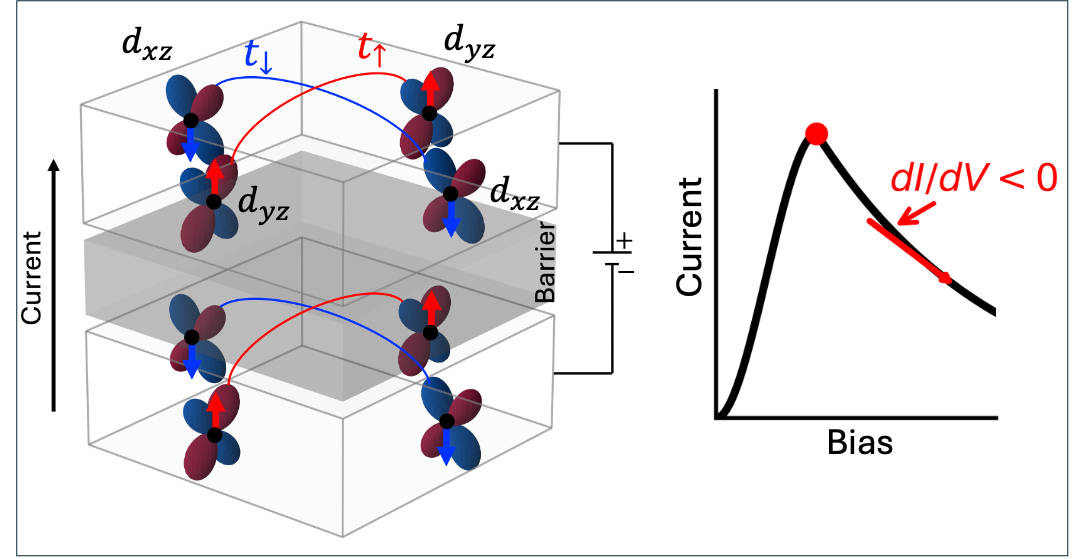}
  
\end{minipage}%

\printbibliography

\end{document}